\documentclass[aps,prd,onecolumn,superscriptaddress]{revtex4}
\usepackage{graphicx}
\usepackage{amsmath}
\usepackage{epsfig}
\usepackage{amsmath,amsfonts,amssymb}
\usepackage{pdfpages}
\usepackage{times}
\usepackage{color}

\begin{document}
\baselineskip=12pt
\def\black{\textcolor{black}}
\def\red{\textcolor{black}}
\def\blue{\textcolor{blue}}
\def\green{\textcolor{black}}
\def\be{\begin{equation}}
\def\ee{\end{equation}}
\def\bea{\begin{eqnarray}}
\def\eea{\end{eqnarray}}
\def\orc{\Omega_{r_c}}
\def\om{\Omega_{\text{m}}}
\def\E{{\rm e}}
\def\bearst{\begin{eqnarray*}}
\def\eearst{\end{eqnarray*}}
\def\peleven{\parbox{11cm}}
\def\peffec{\peight{\bearst\eearst}\hfill\peleven}
\def\pspace{\peight{\bearst\eearst}\hfill}
\def\ptwelve{\parbox{12cm}}
\def\peight{\parbox{8mm}}
%\twocolumn[\hsize\textwidth\columnwidth\hsize\csname@twocolumnfalse\endcsname

\title{Measuring the baryon fraction in cluster of galaxies with \\ Kinematic Sunyaev Zeldovich and a Standard Candle }

\author{Shant Baghram}
\email{baghram-AT-sharif.edu}
\address{Department of Physics, Sharif University of Technology, P.~O.~Box 11155-9161, Tehran, Iran}

\begin{abstract}
We propose a new method to use the kinematic Sunyaev-Zeldovich for measuring the baryon fraction in cluster of galaxies.
In this proposal we need a configuration in which a supernova Type Ia  resides in a brightest cluster galaxy of intermediate redshift clusters.
We show this supernova Type Ia can be used to measure the bulk velocity of a galaxy cluster. We assert that the redshift range of $0.4<z<0.6$ is suitable for this proposal. The  main contribution to the deviation of standard candles distance modulus from cosmological background prediction in this redshift range comes from peculiar velocity of the host galaxy and gravitational lensing. In this work we argue that by the knowledge of the bulk flow of the galaxy cluster and the cosmic microwave background photons temperature change due to kinematic Sunyaev-Zeldovich, we can constrain the baryon fraction of galaxy cluster. The probability of this configuration for clusters is obtained. We estimate in a conservative parameter estimation the large synoptic survey telescope can find spectroscopically  followed $\sim 4500$ galaxy clusters with a bright cluster galaxy which hosts a type Ia Supernova each year. Finally, we show the improving of the distance modulus measurement is the key improvement in future surveys which will be crucial to detect the baryon fraction of cluster with the proposed method.
% PACS numbers:
%04.50.+h, 95.36.+x, 98.80.-k
\end{abstract}

\maketitle

%\newpage

\section{INTRODUCTION}

The baryon density of the Universe and its ratio with respect to the dark matter is fixed by Cosmic Microwave Background (CMB) radiation  \cite{Ade:2015xua} and Big Bang Nucleosynthesis (BBN) to a fraction of $f_b = \rho_b / \rho _m \simeq 0.16$ \cite{Steigman:2007xt}. However this ratio is smaller by a factor of  2 or 3 in galaxies and cluster of galaxies. This is known as the missing baryon problem \cite{Bregman:2007ac}. There is an idea that the missing baryons are in intergalactic medium in diffuse warm-hot plasma which is hard to detect in X-ray. Also there are hints that baryons are not only in the virialized gravitationally bound objects but also they reside in voids and filaments \cite{Eckert:2015akf,deGraaff:2017byg}. Worth to mention that there is an idea that a considerable amount of missing baryon could be in cold transparent molecular clouds, which can be detected via optical scintillation method \cite{Habibi:2013sd}.
The search for the missing baryons is one of the ambitious quests of cosmology, which will shed light on the process of galaxy formation and evolution.  One of the promising cosmological probes to address this problem is the study of the galaxy clusters as the largest gravitationally bound systems which represent the last step of the hierarchical structure formation. It seems that the galaxy clusters suffer less from the missing baryon problem than the other structures and they can be considered as  prominent candidates to check the universality of the baryon fraction ratio in the structures\cite{McGaugh:2009mt}.
We can learn about the physics of baryons in galaxy clusters by studying the interaction of the galaxy cluster with CMB photons. Although less than one percent of the CMB photons passed through the galaxy clusters but the physics of the interaction is known and under control.
The inverse Compton scattering of CMB photons by hot intra-cluster gas of electrons change the intensity of the observed CMB. This effect is known as Sunyaev-Zeldovich (SZ) effect \cite{Zeldovich:1969ff,Sunyaev1972,Birkinshaw:1998qp}.
The bulk motion of the galaxy cluster also introduce a Doppler shift effect on the CMB photons known as kinetic Sunyaev Zeldovich (kSZ) effect \cite{Sunyaev:1980nv}.  The kSZ is a physical process of electron-photon scattering which is color blind (i.e. keeps the CMB spectrum almost unchanged), while the thermal Sunyaev-Zeldovich (tSZ) is a process that changes the CMB spectrum. We should note that for a typical cluster
of galaxies the thermal velocities are higher than the bulk velocity and accordingly kSZ amplitude is an order of magnitude smaller than
the tSZ. Thermal SZ is observed via CMB temperature \cite{Das:2013zf,George:2014oba} and also from individual cluster studies\cite{Plagge:2009eu,:2012vza}. The kSZ is more challenging to be detected because of its smaller amplitude and also the fact that it does not change the spectrum of CMB. Despite to its observational challenges, kSZ is a very valuable quantity to be measured in clusters where it can be used to address some astrophysical questions like  missing baryon problem \cite{Hernandez-Monteagudo:2015cfa,Schaan:2015uaa} and also it can be used to address cosmological questions (e.g. kSZ can be used as a method to measure the growth of structures to constrain dark energy and modified gravity theories \cite{Diaferio:1999ig,Bhattacharya:2006ke,Bhattacharya:2007sk,Bianchini:2015iaa}).
It is worth to mention that there is a frequency band in CMB observations $\nu\sim 218$GHz where the tSZ effect on the intensity of CMB photons is almost zero.
The first detection of kSZ is reported by Hand et al. (2012) \cite{Hand:2012ui} used the correlation of the Atacama Cosmology Telescope (ACT) data \cite{Swetz:2010fy} with the pair-wise velocity of the Baryon Oscillation Spectroscopic Survey (BOSS) spectroscopic catalogue \cite{Ahn:2012fh}. The kSZ signal is also detected in Planck and Sloan data\cite{Ade:2015lza}, South Pole Telescope and Dark Energy Survey cross correlation as well \cite{Soergel:2016mce}.\\
We should note that there are attempts to detect the kSZ signal from individual clusters (for example see \cite{Holzapfel:1997ui,Benson:2003va, Sayers:2013ona, Sayers:2015vha}). The case study of the kSZ signal in individual cluster of galaxies is an important attempt to understand the thermodynamic, merger and evolution of each cluster. \\
In this work we propose a novel idea to measure the baryon fraction in cluster of galaxies by using the kSZ effect. The temperature change due to the kSZ is proportional to the baryon fraction and the bulk velocity. In the case if we can find out the bulk velocity of the galaxy cluster then we can pin down the baryon fraction more accurately by adding the CMB-kSZ information. The main idea of this work is centered on the individual cluster kSZ signal and bulk velocity measurement.  The bulk velocity is usually measured by using the X-ray catalog of clusters as a complimentary probe to kSZ \cite{Mody:2012rh} or it is obtained via reconstructed matter density field \cite{Li:2014mja} or by using the distance indicators\cite{Bahcall:1996bc}. In this direction we suggest to use standard candles such as Supernova (SNe) type Ia to measure the bulk velocity. Traditionally SNe Ia used as a probe of cosmic bulk flow \cite{Appleby:2014kea}. In this work we assume that the SN resides in the brightest cluster galaxy (BCG), accordingly this can be used as a probe of bulk velocity of cluster. Finally adding up the two independent observations of kSZ and SNe Ia, one can use it as a probe of baryon fraction in cluster. We also discuss the observational prospects of this proposal and the probability that this configuration can be observed.
The structure of this work is as follows: In Sec. (\ref{Sec2}) we discuss the theoretical framework of kSZ.
In Sec. (\ref{Sec3}) we discuss the idea of measuring the peculiar velocity via SNe type Ia. In Sec. (\ref{Sec4}) we discuss the observational prospects of the idea raised in this work and finally we conclude in Sec.(\ref{Sec5}).

\section{kSZ and the baryon fraction}
\label{Sec2}
The CMB temperature is changed due to kinetic Sunyaev-Zeldovich (kSZ) effect as
\be
\frac{\Delta T}{\bar{T}}|_{kSZ}(\hat{n})=-\frac{\sigma_T}{c} \int \frac{d\chi}{1+z}e^{-\tau(\chi)}n_e(\chi\hat{n},\chi)\vec{v}_e.\hat{n},
\ee
where $\bar{T}$ is the mean CMB temperature, $\chi= \chi(z)$ is the comoving distance which is a function of redshift, $\sigma_T$ is  Thomson scattering cross section, $\tau$ is the optical depth, $n_e$ is the physical number density of free electrons, $\vec{v}_e$ is the peculiar velocity of free electrons and $\hat{n}$ is the direction defined from observer to source. The optical depth is defined as $\tau = \int dl n_e\sigma_T$, where $l$ is the physical length of the ionized patch in the sky. Note that the optical depth is very small and we set $e^{-\tau} \simeq 1$ hereafter. Also note that the kSZ has almost frequency independent amplitude, for more details see Sec. (\ref{Sec4}).\\
The kSZ signal is an order of magnitude lower than the tSZ %and also we should note that the kSZ does not change the black body spectrum of the CMB.
and mainly is extracted statistically from the cross correlation with the late time tracers of gravitational potential.
In this framework the leading science is done by Atacama Cosmology Telescope(ACT), which detect the kSZ effect by correlating the signal with reconstructed velocity field obtained for the galaxies in LSS surveys such as Baryon Acoustic Oscillation Sky Survey (BOSS)\cite{Schaan:2015uaa}.
%Assuming the baryon fraction in the Universe, one can estimate the optical depth of each galaxy cluster and relate to the kSZ effect
%in a phenomenological relation $\frac{\Delta T}{\bar{T}}|^{i}_{kSZ}=-\alpha\tau^{i}c^{-1}(\vec{v_e}^i.\hat{n})$
%where superscript $i$ indicate the indices for the number of galaxy cluster and $v_e$ is the peculiar velocity of electrons.
However there are attempts to obtain the kSZ signal from an individual cluster (e.g. SZ measurement of MACS J0717.5+3745 galaxy cluster at redshift $z=0.55$  using the data obtained with the NIKA camera at the IRAM 30m telescope \cite{Adam:2016abn} and SZ measurement of RX J1347.5-1145 galaxy cluster at redshift $z=0.45$ with data collected at 147, 213, 281, and 337 GHz using the Multiwavelength Submillimeter Inductance Camera (MUSIC) and at 140 GHz with Bolocam \cite{Sayers:2015vha}). These type of studies are essential to understand the physics of baryons in a galaxy cluster. For example a kSZ signal plus the knowledge on properties of galaxy cluster can be used to obtain the peculiar velocity of free electrons.\\
Now let us look to this problem in another point of view, if we can find the peculiar velocity of the galaxy cluster from an independent way, then we can use the kSZ signal to extract the information about the baryon fraction in  a cluster of galaxies.
One way to reconstruct the peculiar velocity is by linear perturbation theory, the semi-local density contrast of matter perturbation can be used
as a probe of peculiar velocity due to the Euler equation \cite{Padmanabhan:2012hf}. The other method is to use the standard candles like SNe type Ia to extract the information.
We will discuss this method extensively in Sec.(\ref{Sec3}). \\ In what follows we reexpress the kSZ signal in terms of a) the physics of baryons, b) the physics of peculiar velocity.
In order to proceed we write number density of electrons $n_e$ as
\be
n_e\simeq\Delta \bar{n}_e,
\ee
where $\bar{n}_e$ is the cosmic background number density of electrons and $\Delta$ is the electron density contrast which is in order of $\simeq 200$ for a virialized cluster. The background number density of electrons can be expressed in terms of cosmological parameters as
\begin{equation} \label{eq:nbar}
\bar{n}_e=\Upsilon \frac{\rho_g(z)}{\mu_e m_p},
\end{equation}
where $\rho_g$ is the mean gas density in redshift z, $m_p$ is the proton mass and $\mu_e$ is the effective number of electrons per nucleon. Accordingly $\mu_e m_p$ is the mean mass per electron. $\Upsilon$ is the ionization fraction which is defined as $\Upsilon = [1 - Y_p (1- N_{He}/4)]/ (1- Y_p /2)$, where $Y_p$ is the primordial abundance of Helium and $N_{He}$ is the number of ionized electrons corresponding to Helium atoms \footnote{ Helium atoms remain singly ionized until $z \sim3$ and below that, they are thought to be doubly ionized \cite{Shaw:2011sy}.}
Now we can relate the gas density to the baryon fraction of universe and matter density of universe as below
\be \label{eq:rhogas}
\rho_g = \frac{3H_0^2}{ 8\pi G} f_g f_b \Omega_m (1+z)^3 ,
\ee
where $f_g$ is the gas fraction of baryons in a cluster and $f_b$ is our crucial parameter of the study, known as baryon fraction parameter. $\Omega_m$ is the matter density parameter and $H_0$ is the Hubble parameter both defined in present time.
Now by assuming that the evolution of all above parameters inside a cluster is negligible and position independent, the kSZ effect will become
\begin{equation}
\frac{\Delta T}{\bar{T}}|^{i}_{kSZ} \simeq - \sigma_T  \ell^i (\Delta \bar{n}_e)\times \vec{\beta}^i . \hat{n},
\end{equation}
where $\ell^i$ is the size of the cluster labeled by superscript $i$ and $\vec{\beta} \equiv \vec{v}/c$.
%The superscript $i$ indicate a specific cluster labeled by "i". $\delta_e$ is the perturbation in distribution of free electrons in the cluster. This term is important in calculating the non-linear effect known Ostriker-Vishniac effect  \cite{Ostriker:1986fua}.
%However in our calculations this term can be neglected in central region of a galaxy cluster
Now by using Eq.(\ref{eq:nbar}) and Eq.(\ref{eq:rhogas}), it is straightforward to show that the kSZ signal amplitude can be written as
\begin{equation}
\frac{\Delta T}{\bar{T}}|^{i}_{kSZ} \simeq - {\cal{C}}^i f_b  \times  (\vec{\beta}^i . \hat{n}),
%\frac{\Delta T}{\bar{T}}|^{i}_{kSZ} = - \left[{\cal{C}}_i f_b \exp({-{\cal{C}}_i f_b}) \right]\times  (\vec{\beta}^i . \hat{n}),
\end{equation}
where ${\cal{C}}^i$ is a specific parameter for each cluster, which depends on the mean redshift, the physics of intra-cluster medium and the line of sight length of the cluster as
\begin{equation}
{\cal{C}}^i (z_i)= \frac{3H_0^2}{8\pi G } \frac{\sigma_T}{\mu_e m_p} (\Delta f_g \Upsilon \ell^i ) \Omega_m (1+z_i)^3  .
\end{equation}
Now ${\cal{C}}^i$  can be written in the terms of cluster's gas fraction, ionization factor and line of sight length
\be \label{eq:ci}
C^i (z_i) \simeq 4 \times 10^{-3}  \times \Omega_m h^2(1+z_i)^3  f_{g}  \Upsilon  \frac{\ell^i}{ Mpc}.
\ee
Finally we can define a parameter  $x^i= -{\cal{C}}^if_b$  for each cluster  which is related to kSZ signal and the bulk flow of the cluster as
\begin{equation} \label{eq:xex}
x^i \simeq ({\frac{\Delta T}{\bar{T}}|^{i}_{kSZ}}\large)/{(\vec{\beta}^i.\hat{n})} ,
\end{equation}
where the approximation works for $\Omega_m = 0.3$, $h = 0.7$, $\ell^i\simeq 1 Mpc$, $\Upsilon \simeq 1$, $f_g \leq 1$ and $f_b \leq 0.16$.
Accordingly by the knowledge of kSZ signal and the peculiar velocity  we can constrain the physics of free electrons and baryon fraction in a cluster.
%We should note that the baryonic matter of the galaxy clusters is dominated by intra-cluster gas, accordingly we for simplification we assume $f_g\sim 1$.
In the next section we will discuss that how SNe type Ia will be a great candidate in order to calculate the bulk flow parameter.

%%% ****************************   %%%%
%%% ****************************   %%%%
%%% ****************************   %%%%

\section{Standard Candles as a probe of peculiar velocity}
\label{Sec3}

The classical methods for peculiar velocity measurements  from galaxy surveys needs an independent distance indicator to subtract the Hubble flow from redshift position ($v_p\simeq cz - Hr_{phy}$ , where $v_p$ is the peculiar velocity, $z$ is the redshift of the galaxy/cluster and $r_{phy}$ is the physical distance.) Many distance indicators are used to measure the peculiar velocity such as Tully-Fisher relation \cite{Tully:1977fu}, fundamental plane \cite{Djorgovski:1987vx} and the SNe Ia as well.
The Tully-Fisher and fundamental plane measurements introduce a $20\%-25\%$  error in peculiar velocity measurement \cite{Strauss:1995fz,Willick:1996cb}.
Although there is an assertion by Springob et al.\cite{Springob:2014qja}  that the fundamental plane  analysis  results to relative error in peculiar velocity  less than $5\%$ at all redshifts. However there is an approximation in \cite{Springob:2014qja}, that the ratio of the apparent and physical size of a galaxy in redshift $z$ is proportional to $(z-z_p)/z$ (where $z_p$ is the redshift assigned to peculiar velocity). This assumption means that in small redshifts the physical distance to the galaxy is equal to $cz/H_0$.
The SNe Ia as a distance indicator is also introduce an error in velocity measurement due the fact that the  peculiar velocity of the host galaxy change the magnitude of the SN Ia \cite{Habibi:2014cva}. The other method that can be used for peculiar velocity measurements is via reconstruction of density field and peculiar velocity\cite{Zaroubi:2002hh}.  We argue that in future experiments the method we propose in this work will be comparable with classic methods of peculiar velocity measurement.
In this section we will discuss how standard candles can be used to determine the peculiar velocity of the structures without the use of standard relation (i.e. $v_p\simeq cz - Hr_{phy}$).
The idea is straightforward, the standard candles are used to establish a luminosity distance-redshift relation for a given background cosmology model.
However the deviation from the homogenous background will change this relation. One of the important modifications is due to the peculiar velocity of the host galaxy of a SN, which affects both the luminosity and redshift of the standard candle.
 Accordingly, we can use the deviation of the luminosity distance of a SN type Ia as a probe of  peculiar velocity. \\
We assume a perturbed FRW universe with a Newtonian comoving gauge metric
\be
ds^2=a^2(\eta)\left[-(1+2\Psi (x,t))d\eta^2 + (1-2\Phi(x,t))\delta_{ij}dx^idx^j\right],
\ee
where $\eta$ is the conformal time, $\Psi$ and $\Phi$ are the scalar perturbations of metric. If we assume that the General Relativity (GR) is the correct classical theory of gravity and the universe is filled with components that have no anisotropic pressure, then  we get $\Phi=\Psi$. \\
In a perturbed universe, the luminosity distance of a standard candle is modulated due to propagation of light in perturbed FRW universe (Sachs Wolfe effect and gravitational lensing). The luminosity distance is also corrected due to the peculiar velocity of the source and observer. These corrections can be formulated as\cite{Bonvin:2005ps,Bacon:2014uja,Baghram:2014qja}
\be
d_L(z_s,\hat{n})=(1+z_s)\chi(z_s)\left[1- \kappa_v-\kappa_g - \kappa_{SW}- \kappa_{ISW}\right],
\ee
where $d_L(z_s,\hat{n})$ is the luminosity distance of a supernova in observed redshift of the source $z_s$ and direction $\hat{n}$ (Note that $\hat{n}$ is unit vector in the direction of observer toward source). $\chi(z_s)$ is the comoving distance of the source in FRW universe. The parameter $\kappa_v$ is the correction due to peculiar velocity of source. The $\kappa_g$ , $\kappa_{SW}$  and $\kappa_{ISW}$ are the lensing convergence, Sachs-Wolfe and Integrated Sachs-Wolfe correction terms. (These terms are defined and studied extensively in \cite{Bacon:2014uja}).
%In the redshift range of $z < 0.15$ the most contribution to the luminosity distance change comes from the peculiar velocity.
In intermediate redshift $0.4<z<0.6$ which is the redshift range which we are interested in this work the peculiar velocity and gravitational lensing has comparable effect, accordingly  \cite{Habibi:2014cva}
\be \label{eq:dlkappav}
d_L(z_s,\hat{n})\simeq\bar{d}_L(z_s)[1-\kappa_v - \kappa_g],
\ee
where $\bar{d}_L(z_s)$ is the background luminosity distance and $\kappa_v$ is luminosity correction due to peculiar velocity  \cite{Bonvin:2005ps} defined as
\be \label{eq:kappav}
\kappa_v= - \left(1- \frac{1+z_s}{\chi(z_s)c^{-1} H(z_s)}\right)(\vec{\beta}_s.\hat{n}),
\ee
where $H$ is the Hubble parameter and $\vec{\beta}_s = {\vec{v}}_p / c$ where ($v_p$ is the peculiar velocity).
In low and intermediate redshifts ($z < 1.4$), the term in parentheses is negative, accordingly the objects moving toward us ($\vec{\beta}_s.\hat{n}<0$) introduce  a $\kappa_v<0$ where if we replace this in Eq.(\ref{eq:dlkappav}) we will get a dimmer SN. In the other hand when the host of a standard candle is moving away from us ($\vec{\beta}_s.\hat{n}>0$), that introduces a $\kappa_v >0$ and accordingly the source become brighter. These chain of conclusions are changed in higher redshifts.
The lensing correction term is defined as
\be
\kappa_g = \int_0^{\chi(z_s)}d\chi(\chi(z_s)-\chi)\frac{\chi}{\chi(z_s)}\nabla_{\perp}^2\Phi\simeq \frac{3}{2}\Omega_m H^2_0\int_0^{\chi(z_s)}d\chi (\chi(z_s)-\chi)\frac{\chi}{\chi(z_s)} (1+z)\delta_m ,
\ee
where $\delta_m$ is the total matter density contrast. Note that the second term is an approximation as we change the 2D Laplacian with a 3D.
%%//////////////////////////////////////////////////////////////////%%
\begin{figure}[ht!]
\centering
\includegraphics[width=0.52\textwidth,natwidth=610,natheight=400]{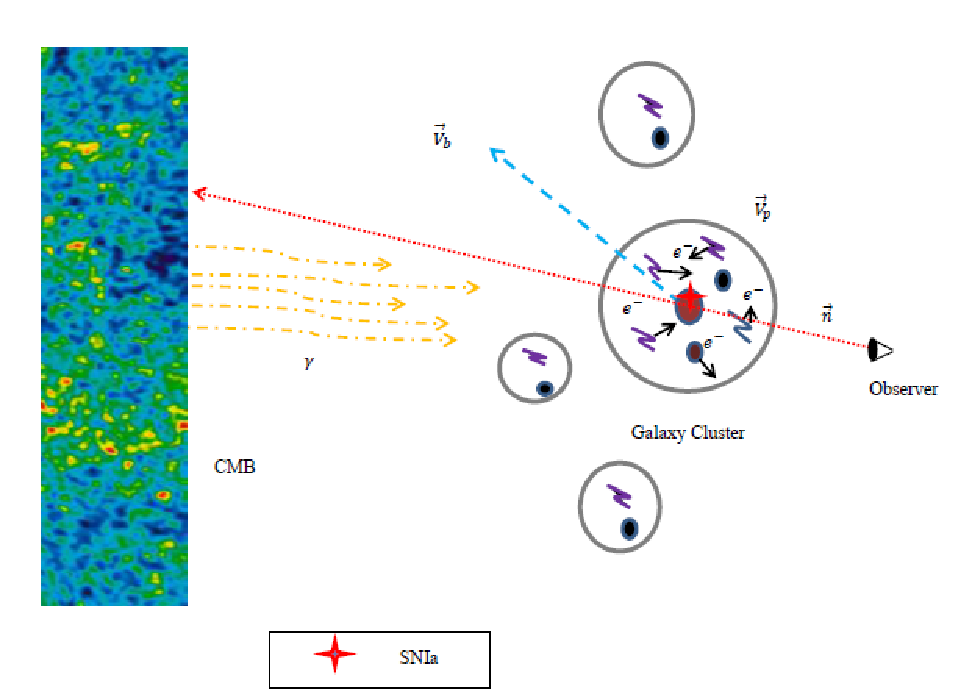}
\caption{A schematic configuration of a galaxy cluster is plotted which has a kSZ signal and its  bright cluster galaxy hosts a supernova type Ia. The dotted red arrow indicate the line of sight direction and the blue dashed arrow is the direction of the bulk flow. The solid black arrows shows the peculiar velocities of cluster members. Orange dashed dotted arrows shows the CMB photons which will interact with ionized plasma of galaxy cluster and the red star represent a SNIa exploded in the BCG.  \label{fig:config}}
\end{figure}
%%//////////////////////////////////////////////////////////////////%%

The main idea of this work is that we can use SNe for finding the peculiar velocity (bulk flow) of the galaxy cluster.
Before proceed further, we should numerate the velocity contributions to our specific case of study. The velocities which can be assigned to a SN Ia are: \\
~~~a) The peculiar velocity introduced from the progenitor of the SNIa, b) The peculiar velocity of the host galaxy due to the gravitational potential of the cluster and c) The peculiar velocity due to the bulk flow of the galaxy cluster.
In this work we are merely interested in bulk flow, accordingly in order to extract this term, which is the source of the kSZ effect, we assume, that the host galaxy of the supernova is the one which is located in the gravitational/ optical center of the galaxy cluster (bright cluster galaxy (BCG)).
The assumption that the SN host galaxy is in the center of the cluster is needed, as we want to assign the bulk velocity of the cluster to the peculiar velocity of the BCG. This assumption comes from the idea that the central bright galaxies reside in the minimum of the gravitational potential well of the cluster and the peculiar velocity with respect to the center of the mass of a cluster is smaller than the total bulk motion of the whole system.\\
We assert that main contribution to the luminosity change of a SN in BCG is due to the peculiar velocity of the bulk motion of the cluster. This configuration is schematically shown in Fig.(\ref{fig:config}). However we should keep in mind that the effect of SNe Ia's progenitor's velocity {\it{may}} have a very significant velocity offset with respect to the peculiar velocity of the BCG host galaxy. This offset is included as a source of an error in our upcoming estimations. %Keeping this in mind we proceed by assigning the velocity corrections of luminosity distance to the bulk flow.
This can be done by the idea that the progenitor velocity introduce a Gaussian error to the velocity estimation  related to the dynamical mass of BCG. The contribution of this error can be calculated by setting the velocity of the center of mass of a SNIa progenitor equal to the dispersion velocity inside a typical BCG using the fundamental plane relation of elliptical galaxies \cite{BoylanKolchin:2006rs,Bernardi:2006qm}, for a specific example, we express the dispersion velocity in terms of magnitude \cite{Lauer:2014pfa} as $log_{10}({\sigma_{pro}}/{300 kms^{-1}}) = - (0.275 \pm 0.023) ({M_m}/{2.5}) - 2.55\pm 0.21$ where $M_m$ is the metric magnitude \footnote{The metric magnitude obtained from measuring the  Luminosity from the central region of BCG which is much smaller than extent of galaxy. The radius is obtained by the correlation of the Luminosity with structural parameter of BCG. For details see \cite{Postman1995}} and $\sigma_{pro}$ is the velocity of the center of mass of a SNIa progenitor. In this work we set the dispersion velocity (progenitor velocity) to $ \sigma_{pro} \sim 300 km/s \pm 10 \%$. \\
%\be
%\frac{\sigma_e}{km/s} \simeq 660  \times [ c_e \times(\frac{M_{dyn}(<R_e)}{10^{13} M_{\odot}}) (\frac{ 100 kpc }{ R_{e}}) ]^{1/2}
%\ee
%where $R_{e}$ is the half light radius of the BCG and $c_e$ is structure coefficient, which encapsulates in it the complexity of galaxy properties and its environmental effects.
%However this is a very crude approximation for the velocity dispersion in BCG. The idea of using the fundamental plane in the 3 dimension of (half light radius $R_e$, dispersion velocity $\sigma$ and
Now by using Eqs.(\ref{eq:dlkappav} and \ref{eq:kappav}), we can find the line of sight normalized velocity as
\be \label{vlos}
(\vec{\beta}_s.\hat{n}) = (1- 10^ {\Delta\mu /5} - \bar{\kappa}_g) / \tilde{\kappa}_v,
\ee
where $\Delta \mu$ is the difference of the observed distance modulus and the one predicted from background cosmology of $\Lambda$CDM in the specific redshift of $z_s$. The parameter $\tilde{\kappa}_v =(1+z_s) cH_0^{-1}/ \chi(z_s) E(z_s) - 1$ is a unique term which is independent of the local physics, instead it depends on the background cosmological parameters and the redshift of the supernova. Note that $E(z_s)$ is the normalized Hubble parameter to its present value.
Note that $\bar{\kappa}_g$ is the correction term due to gravitational lensing term, and the bar indicate that this term can be obtained by using the matter power spectrum of density perturbations.
\be \label{eq:kg}
\bar{\kappa}^2_g = \frac{9}{4}\Omega^2_m H^4_0\int_0^{\chi(z_s)}d\chi \left[ (\chi(z_s)-\chi)\frac{\chi}{\chi(z_s)} (1+z) \right]^2 \int_0^{\infty}\frac{k dk}{2\pi}P_m(k,z) ,
\ee
where $P_m(k,z)$ is the matter power spectrum in redshift $z$. Now by using Eq.(\ref{vlos}) we can calculate the line of sight velocity of a SN, then we can assign this to the bulk velocity of the galaxy cluster which is the host of the SN $\vec{\beta}_s.\hat{n} \equiv (\vec{\beta}^i.\hat{n}) $ where superscript $i$ represent the host of a SN. Now  we can use Eq.(\ref{eq:xex}) to extract the baryon fraction. This can be done because the RHS of this equation is fixed by two independent observations.
As mentioned in introduction and the beginning of this section the peculiar velocity can be measured by different method like velocity reconstruction with a galaxy field\cite{Ho:2009iw} or using distance indicators, which can be cross-checked with other methods. Now that we obtained the peculiar velocity via SNe Ia method, we should mention the peculiar velocity of the the host introduce an error on its measurement which is proportional to $\delta v_p / v_p \propto H(z_s)\bar{r}_{phy} \kappa_v  / (cz_s - H \bar{r}_{phy})$, which this new source of uncertainty is not considered in classical methods of peculiar velocity measurement by distance indicators.
In the next section we will discuss the observational prospects of finding the missing baryons by the method described in this work.

%%//////////////////////////////////////////////////////////////////%%
%%//////////////////////////////////////////////////////////////////%%

\section{Observational Prospects}
\label{Sec4}

In this section, we will discuss the observational prospects of the idea proposed in previous sections.
In the first subsection, we use the data sample of Union 2.1 in order to represent a logical path of extracting the bulk flow of the clusters from the data with assuming that each SNIa resides in a BCG. We should note that the first subsection is just an example of how this method works and shows the current status of quality of the data. In the second subsection we discuss the error estimation for realistic and optimistic case for finding the baryon fraction and finally in the third subsection we present an estimate for the number of events that we anticipate in future surveys which are suitable for our case of study.
\subsection{Union 2.1 SNe Ia data sample as preliminary example}
In this direction first we assume that all the SNe data from the known catalog are a potentially plausible candidates for  our proposal. We assume that they are hosted by a central galaxy of a cluster. This is just an assumption to show the details of the proposal.
In this section, we use the Union 2.1 SNe sample \cite{Suzuki:2011hu} to show the procedure.
First of all we  extract the peculiar velocity with the method which is described in Sec.(\ref{Sec3}) by assuming that the standard $\Lambda$CDM model with best parameters fixed by supernova data \cite{Ade:2015xua} describe the cosmological flat background ($\Omega_m =0.297$, $h = 0.704$). We obtain the difference of distance modulus $\Delta \mu = \mu_{obs} - \bar{\mu}$ versus redshift for the sample of SNe. Note that $\bar{\mu}$ is the distance modulus from the known background cosmology.

\begin{figure}
\centering
\includegraphics[width=0.5\textwidth]{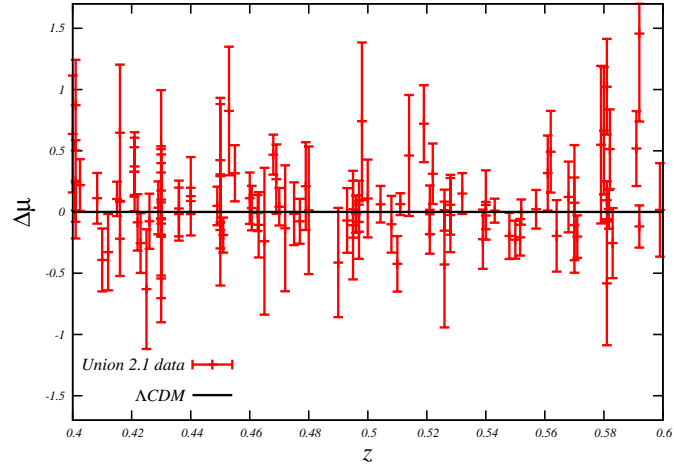}
\caption{The difference of the SNe Ia distance modules with respect to the $\Lambda$CDM prediction ($\Delta \mu$) is plotted versus redshift  for Union 2.1 in the redshift range of $(0.4 , 0.6)$ with $1\sigma$ error bar. The black solid zero line shows the baseline of $\Lambda$CDM for comparison.}\label{fig:deltamu}
%Low amplitude dashed blue curves are obtained from linear theory and the high amplitude green dotted curves are set by assuming a maximum velocity of $1200 km/s$ for a typical cluster.}\label{fig:deltamu}
\end{figure}
%% ****

\begin{figure} \label{fig:deltamu1}
\centering
\includegraphics[width=0.5\textwidth]{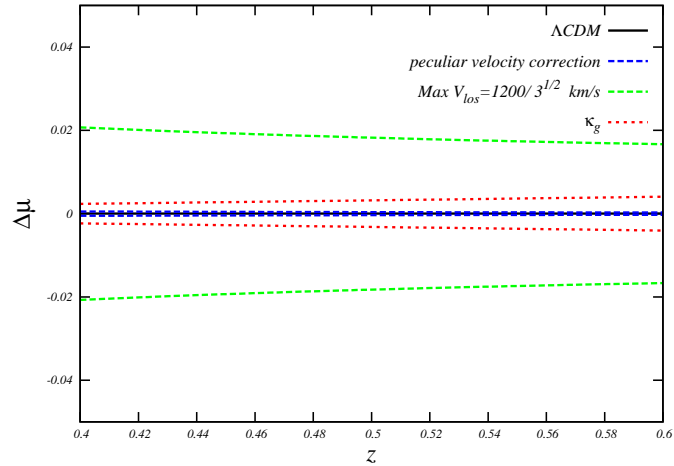}
\caption{The theoretical difference of the SNe Ia distance modules with respect to the $\Lambda$CDM prediction ($\Delta \mu$) in the perturbation level is plotted versus redshift. The dashed blue curves are obtained from linear theory prediction for peculiar velocity corrections and the green long dashed lines are for velocity corrections with larger amplitude, assuming a maximum velocity of $1200 km/s$ for a typical cluster. The red dotted line correspond to the correction to the magnitude due gravitational lensing convergence.}\label{fig:deltamu}
\end{figure}

In Fig.(\ref{fig:deltamu}) we plot $\Delta\mu$ versus redshift for SN Ia in redshift range of $(0.4 , 0.6)$. In this redshift range the host galaxy cluster of the SN can be observed in arc-minutes. This resolution is needed for kSZ measurements. The clusters in lower redshift are observed in lower moments (large angular resolution), where the primordial CMB dominate the kSZ signal. On the other hand in higher redshifts the peculiar velocity corrections to SNe data become subdominant plus the fact that the errors in SN magnitude become larger. % As it is mentioned, in this redshift range the peculiar velocity correction is the dominant effect in the luminosity change of the SNe Ia \cite{Bacon:2014uja,Habibi:2014cva}, where we neglect the contribution of weak lensing convergence and the Sachs-Wolfe effect.
In Fig.(\ref{fig:deltamu}) the  red data points are $101$  SNe Ia in the redshift rang of our interest with average error of $\sim 0.304 + \pm 0.081$ mag.
In Fig.(\ref{fig:deltamu1}), we plot the corrections we anticipate from peculiar velocity in linear theory (blue long dashed line) and to the typical peculiar velocity of $v_{p} \sim 1200 / \sqrt{3} km/s $. The red line is the contribution due to gravitational lensing which is considered as a source of correction to the distance modulus. The convergence corrections is obtained from Eq.\ref{eq:kg}. From the plot it is obvious that the lensing correction increase with the redshift while the peculiar velocity corrections decrease. We choose the redshift range in a way that we can observe the cluster of galaxies with CMB experiments and extract their kSZ effect and also be in a plausible range where the corrections to distance modulus due to peculiar velocity are calculable.

The blue low amplitude dashed curves represent the amount of correction that we expect from the peculiar velocity in linear regime. The linear regime velocity in the line of sight can be obtained via linear matter spectrum $P(k)$ as:
\be
\langle v^2_p \rangle (z) \simeq \frac{1}{3} \frac{H_0^2 f^2}{2\pi^2} \int P(k)W^2(kR(z))dk,
\ee
where $\langle v^2_p \rangle (z)$ is the average linear velocity in the line of sight (assuming an isotropic velocity) in a window function with a comoving radius $R$. The parameter $f$ is the growth rate which for standard $\Lambda$CDM is equal to $f\simeq [(\Omega_m (1+z)^3) / E(z) ]^ {0.55}$ (Note that $E(z)$ is the normalized Hubble parameter to its present value). The blue curves shows the prediction of standard model in perturbation level. The dotted high amplitude green lines are obtained by assuming a maximum line of sight velocity of $\sim 1200 / \sqrt{3} km/s \sim 693 km/s$.
%The black points with their $1\sigma$ error-bars indicate that the chosen SNe Ia have a tension with background prediction as $\Delta\mu$ is not zero.
%However they are consistent by linear perturbation prediction or with a maximum velocity of $1200 km/s$ for a typical cluster. Accordingly we neglect the SNe that  are out of the region of green curves with considering their error-bars (This set of SNe are also plotted in red points).
%The tension with the background can be interpreted as the effect of the first order correction to the distance modulus due the fact that cosmological model deviates from homogenous-isotropic background. Accordingly the black data points of SNe, potentially, can be interesting candidates in order to use them to extract the line of sight velocity of their host galaxies. In this procedure we fix the background parameters by merely using the SNe Ia data, however we can fix the background cosmology by complimentary observations such as CMB temperature anisotropy power spectrum. It is worth to mention that $58$ SNe Ia data pass the mentioned criteria.
In the next step, we assign the deviation of the distance modulus change to the peculiar velocity of the host galaxies.

\begin{figure}
%\centering
\includegraphics[width=0.5\textwidth]{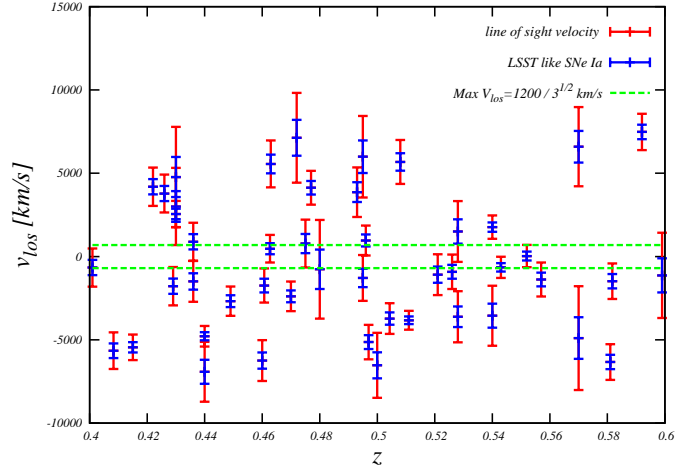}
\caption{The line of sight velocity of SNe is obtained from Eq.(\ref{vlos}) is plotted versus redshift for the SNe of the Union sample. The horizontal dashed green lines shows the typical maximum velocity of galaxy clusters $v_p\simeq \pm 1200 km/s$. This plot is obtained for realistic and optimistic errors on SNe luminosity distance measurement. The uncertainty due to the progenitors are the same in considered cases. Note that the SNe Ia have less than $2\sigma$ tension with the prediction of  distance modulus change via $v_p\simeq \pm 1200 km/s$ peculiar velocity}
\label{fig:betta}
\end{figure}

\begin{figure}
%\centering
\includegraphics[width=0.5\textwidth]{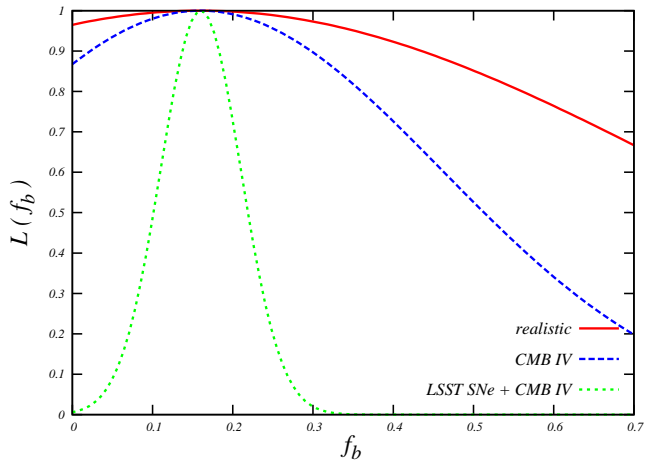}
\caption{The $68\%$ constraints on fiducial parameter of baryon fraction due to realistic (current) errors on physics of cluster, kSZ and SNe Ia luminosity distance measurement and the optimistic errors for CMB stage IV (blue dashed line) and CMB stage IV and LSST like SNe Ia data (green dotted line)}
\label{fig:fisher}
\end{figure}

In Fig.(\ref{fig:betta}) we plot the reconstructed line of sight velocity in terms of redshift. We use Eq.(\ref{vlos}) to extract the line of sight velocity. The error bars are obtained from the propagation of the error in distance modulus and gravitational lensing corrections and in this step also an error is added due to the velocity of the SNe progenitors. For this task we use a Monte-Carlo method to add a Gaussian error $\sigma_{pro} \sim 300 km/s \pm 10\%$.
The very interesting point to indicate here is that the SNe that seems brighter ( $\Delta\mu<0$) moves away from us $\vec{\beta}.\hat{n}>0$. This observation is consistent to the argument we made before for peculiar velocity effect in low and intermediate redshifts.
In Fig.(\ref{fig:betta}), we also obtained the peculiar velocity with an optimistic resolution. The data with smaller blue error-bars come from the assumption that the error-bars of distance modulus in the optimistic case become $\sim 0.2$ mag in average ($30\%$ improvement), which is the characteristic magnitude resolution which can be obtained from LSST survey \cite{Abell:2009aa}.  Fig.(\ref{fig:betta}) shows that how the accuracy in SNIa luminosity distance measurement are important in this study. We should also note that the SNe Ia line of sight velocity is valuable for each cluster and the binning of data like the case of bulk velocity measurement is not useful.   \\
The baryon fraction can be obtained by the knowledge of kSZ signal for each galaxy cluster and the assumption that the SNIa  host is the BCG as below
\be \label{eq:fb}
{
{f^i_b \simeq \left[ (\frac{\Delta T}{\bar{T}}|^{i}_{kSZ}) / {\left(10 ^ {\Delta \mu /5} -1 - \bar{\kappa}_g\right)}\right] \frac{\tilde {\kappa_v}}{C^i}}},
\ee
where $C^i$ obtained from Eq.(\ref{eq:ci}) and $\tilde{\kappa}_v$ depends on the background cosmology. We should note that Eq.(\ref{eq:fb}) is the key equation of our proposal. The baryon fraction of cluster can be obtained by its kSZ temperature and the luminosity change of a SN which is hosted by the central galaxy of a cluster.
In the next subsection we discuss the error propagation due to other components of Eq. (\ref{eq:fb}), which will introduce the uncertainty in baryon fraction calculation.
In Fig. (\ref{fig:fisher}), we plot the Fisher forecast of the baryon fraction measurement via SNe Ia in the redshift range of $(0.4,0.6)$ for our configuration. We set the fiducial parameter $f_b = 0.16 $ and plot the $1\sigma$ prediction of our method for realistic and optimistic cases. In the optimistic case we consider two categories of a) The CMB stage IV case with $S/N \sim 10$ for kSZ measurement. (dashed blue line) and  b) The CMB IV + LSST like SNe (dotted green line). The optimistic cases in this stage is just chosen to show how the baryon fraction can be constrained by future experiments. %Note that $\sigma_{sn}$ is the error due to the SNe Ia uncertainties related to intrinsic magnification errors, photometric errors and etc.
In the next subsection, we discuss the error estimation and we study in more detail how different aspects of this proposal introduce uncertainties. Also we will show that the improvement in luminosity distance measurement has the main contribution for make this proposal a observationally practical one.

\subsection{The error estimation}
In this section, we study the different error budget which has a role in baryon fraction measurement via the method which is proposed in this work.
For simplicity we assume that the background cosmological model is fixed by other observations such as CMB and the errors on density parameters and Hubble constant is much more smaller than the uncertainties in the physics of the cluster, supernova observations and kSZ signal extraction. This means that the uncertainty in baryon fraction calculation can be written as
\be
\sigma^2_{f_{b}} = (\frac{\partial f_b}{\partial (\Delta T / T |_{kSZ})})^2\sigma_{ksz}^2 + (\frac{\partial f_b}{\partial \Delta\mu})^2 \sigma^2_{sn} + (\frac{\partial f_b}{\partial C_i})^2\sigma^2_c + (\frac{\partial f_b}{\partial \bar{\kappa}_g})^2\sigma^2_{\kappa_g},
\ee
where $\sigma_{ksz}$ is the error in kSZ signal from CMB analysis $\sigma_{sn}$ is due to SNe Ia uncertainties which can be due to intrinsic magnification errors, photometric errors, the peculiar velocity of the SNe Ia progenitors and ...  $\sigma_c$  is the error induced from the physics of the clusters and $\sigma_{\kappa_g}$ is the error due to gravitational lensing effect. We also assume that errors from different contributions are uncorrelated and Gaussian distributed, which is almost a reasonable assumption. In what follows we will numerate the different contributions to the total error introduced in this proposal and then we will address that which errors can be reduced in future experiments. \\ \\
{\it{Kinematic Sunyeav Zeldovich}}: For kSZ the main problem is that the spectrum has almost a flat power in frequency range, that is why the primordial CMB anisotropies themselves are the most  important source of contamination. Accordingly most of kSZ extraction methods try to separate the  cluster signal of kSZ and primordial anisotropies.
In the case if the galaxy cluster can be observed in redshifts high enough to be studied in arc-minute resolution then we can use the kSZ measurements. The arc-minute resolution  corresponds to angular moment of $\ell \sim 3000$, the CMB primordial signal become less important due to Silk damping and accordingly the kSZ signal can be extracted from primordial one. %However the method we proposed here works for galaxy clusters $z<0.15$, in this case the kSZ signal extraction become a difficult task.
%Regarding the angular resolution, the galaxy clusters which is observed in order of a few arc-minutes resolution (angular moment of $\ell \sim 3000$) are in high redshift $z\sim 1$ .
The other source of contamination is the tSZ effect, which is an order of magnitude higher than the kSZ in ordinary clusters.
An important step toward separation of kSZ from tSZ is multi-frequency observations of individual clusters. The total intensity change $\Delta I_{\nu}$ due to a galaxy cluster observed in frequency $\nu$ with respect to CMB $I_0$ is given as
\begin{equation}
\frac{\Delta I_{\nu}}{I_0} = f(\nu,T_e)y_{tsz}+g(\nu,T_e,v_p)y_{ksz},
\end{equation}
where $T_e$ is the temperature of intra cluster medium (ICM) , $v_p$ is the line of sight velocity. $y_{tsz}$ and $y_{ksz}$ are the amplitude of tSZ and kSZ signal and $f(\nu,T_e)$ and $g(\nu,T_e,v_p)$ are the characteristic frequency dependent function of tSZ and kSZ as \cite{Birkinshaw:1998qp}
\begin{equation}
f(x,T_e)= \frac{x^4e^x}{(e^x-1)^2}(x\coth(\frac{x}{2})-4)(1+\delta_{tsz}(x,T_e)),
\end{equation}
where $x=h\nu/k_BT_{CMB}$ and $\delta_{tsz}(x,T_e)$ is the relativistic correction to the characteristic function\cite{Itoh:2003dp}.
\begin{equation}
g(x,v_p,T_e) = \frac{x^4e^x}{(e^x-1)^2}(1+\delta_{ksz}(x,v_p,T_e)),
\end{equation}
where  $\delta_{ksz}(x,v_p,T_e)$ is the relativistic correction to the characteristic function\cite{Nozawa:2005ut}.
Now if we have signal in two frequency bands (like The New IRAM KID Array (NIKA) camera at the IRAM 30m telescope at $\nu_1=150$ and $\nu_2=260$ GHz \cite{Catalano:2014nml} we can extract the amplitude of kSZ amplitude as below
\begin{equation}
y_{ksz}=\frac{f(x_1,T_e)\Delta I_{\nu_2} - f(x_2,T_e)\Delta I_{\nu_1} }{I_0 f(x_1,T_e) g(x_2,v_p,T_e) - I_0 f(x_2,T_e) g(x_1,v_p,T_e)}
\end{equation}
where $x_i=h\nu_i/k_BT_{CMB}$ , $i=1,2$. This procedure is done for observing the cluster MACS J0717.5+3745 with NIKA  which is composed of four distinguishable sub-clusters. The detection of kSZ signal in one of sub-clusters is obtained by $4.6\sigma$ \cite{Adam:2016abn}. Also the technique of component separation is used in a recent work to extract the kSZ signal to study the thermodynamic of a cluster \cite{Battaglia:2017neq}. One of the main future project which will use the technique of multi-wavelength detection of SZ effect is the 6-meter CCAT-prime telescope\cite{Mittal:2017hwf}.
Note that in distinguishing procedure, the X-ray observation of the clusters can be considered essential and also complimentary in order to study the effective temperature of ICM \cite{Adam:2016abn}.
There is other multi-wavelength studies of the MACS J0717.5+3745 such as the study of the SZ with MUSTANG and Bolocam \cite{Mroczkowski:2012ew}
Except the primordial CMB anisotropy and tSZ, there are other sources of noise such as diffuse foreground emissions which can be divided to atmospheric, galactic and extragalactic sources.
{ For comparison on the level of error dependence, we set the relative uncertainty as $\sigma_{ksz} / (\Delta T / T |_{kSZ}) \simeq  100\%$  (i.e. $S/N \sim 1$) for a realistic case \cite{Swetz:2010fy, Alonso:2016jpy} and for futuristic (e.g. CMB stage IV experiments) one we set $\tilde{\sigma}_{ksz (opt)} \simeq 0.10  \tilde{ \sigma} _{ksz (r)} $  (i.e. $S/N \sim 10$) \cite{Alonso:2016jpy}.  (Note that tilde is an indication of  relative errors and subscripts "r" and "opt" are for realistic and optimistic cases respectively).}  \\ \\
%In the case of SNe Ia magnitude errors, we have two contributions: one is from the intrinsic error in SNIa magnification and the other is from the progenitor of SNIa accordingly the relative uncertainty of SNIa become $ \sigma_{sn} / \Delta \mu   = [ (\sigma_{sn(i)} / \Delta \mu)^2 + (\sigma_{sn(p)} / \Delta \mu)^2]^{1/2}$. For the realistic case the the intrinsic relative error is obtained by setting an accuracy of one tenth of magnitude and for futuristic experiment like LSST we set the relative intrinsic error  to $\tilde{\sigma}_{sn (opt)} = 0.5 \tilde{\sigma} _{sn(r)}$. For both cases the progenitor relative error keep fixed.
%As the uncertainty in the SNe Ia has the main contribution to the baryon fraction, we present our forecast with $\tilde{\sigma}_{sn (opt)} = 0.1 \tilde{\sigma} _{sn(r)}$ as well.
{\it{Supernovae Type Ia}}: In the case of the SNe Ia, nowadays with the observation of around $10^3$ supernovae the systematics become the dominant one in error budget in comparison to the statistical errors \cite{Howell:2010vd}. This systematics are also survey dependent, accordingly all the supernovae compilation have the problem of calibration matching where photometric offsets is one of the main challenges of using more than one survey. In this direction the future surveys like LSST which has the plan to make a dedicated long run surveys of sky for hunting the SNe Ia, can overcome this problem \cite{Abell:2009aa}. Generally the systematics of the SNe Ia intrinsic luminosity measurement can categorized due to the effects such as:\\ a) {\it{calibration}}: one of the major obstacles in low redshift SNe Ia measurements becomes from the usage of old Landolt photometric system. However for each SN we need a calibration by knowing the filter transmission rate and K-correction. Note that one needs the knowledge of the spectral energy distribution (SED) for K-correction \cite{Hsiao:2007pv}. Also we should indicate that large number of low redshift SNe will solve this problem by using a new system of calibration.\\  b) {\it{the ultraviolet treatment of high redshift SNe Ia}}: one of the main challenges of high redshift $z>0.2$ SN is the intrinsic scatter in their light curves and their poor calibration \cite{Kessler:2009ys}.\\ % In the case of our proposal, we can neglect this error as we are dealing with low redshift SNe Ia.\\
c) {\it{reddening due to dust}}: This is one of the major uncertainties introduced in the modeling of the SNe Ia. It seems that the redder SNe are also dimmer \cite{Riess:1996pa}. This means that there is an intrinsic color-luminosity relation in SN data which can become degenerate from the reddening of the host galaxy's dust. As the color is correlated to the SN's properties so associating the wrong color to the host of SN can be wrongly assigned to the reddening. The idea to overcome this error is to study the physics of the color-luminosity and reddening simultaneously with independent observations. The infrared observations of the host galaxy can be a solution to decrease the contamination of the dust \cite{Shariff:2016cxw}.  A final note in this category is that the different SN progenitors can have different circumstellar dust properties.\\ d) {\it{environment dependence and redshift evolution of SNe Ia}}: It seems that there is an indication that most luminous SNe Ia occurred in late type galaxies \cite{Hamuy:1996sq} and sub luminous ones are in old population galaxies \cite{Howell:2001hr}. This can be regarded as a strong evidence for redshift and environment dependence of SNe Ia. Now we have an indication that the width and luminosity of SNe Ia light curve is correlated to the mass, star formation rate and the metallicity of the host galaxy \cite{Sullivan:2006ah,Gallagher:2008zi,Sullivan:2010mg}. As the star formation rate is a redshift dependent quantity accordingly we can anticipate that the SNe Ia absolute magnitude can be a redshift dependent as well. However, now we are entering a large data set era, which by comparing the SNe Ia and their host galaxies properties we can improve the SNe Ia modeling. In other words the large statistics can be used to control the systematics. \\ e) {\it{the physics of progenitor}}: the last but not the least effect of the error estimation of SNe Ia as a standard candle is the progenitor. There are mainly three scenarios of SNe Type Ia,  known as single degenerate (SD) scenario \cite{Whelan1973}, double degenerate (DD) scenario  \cite{Iben1984,Webbink1984} and  sub-Chandra \cite{Weesley1986}.  The first scenario assumes that a white dwarf with a companion from main sequence of stars or a red giant is the generator of the nova and in the second one, we assume that the binaries are both white dwarfs. The  sub-Chandra model assumes that a layer of helium appears  on the surface of a white dwarf below the Chandrasekhar mass until it detonates.
The main goal of future SNe Ia survey like LSST is to observe a large number of SNe Ia (e.g. 50,000 Type Ia per year in LSST survey ), where the good statistics in color and light curves, combined with a  small number  of sample spectra,  any dependence of the supernova standard candle relation on parameters
other than light curve shape and extinction can be extracted. Accordingly these can be minimize the systematics via statistics. The LSST has a plan to reduce the distance indicator error to $0.2$ mag which independently can constrain the dark energy equation of state with statistics better than $10\%$.
\\
Another important observation which can shed light on the distance measurements in local universe is the GAIA project which is a space observatory of the European Space Agency (ESA) designed for astrometry which can be a great help to calibrate the  cepheid variables and make the distance ladder more accurate \cite{Barstow:2014dda}. \\ \\
{\it{Galaxy cluster}}: The  main contribution to the relative error of the physics of galaxy cluster is raised form the size of the galaxy and the gas fraction of the cluster. The relative uncertainty is set $\tilde{\sigma}_c \simeq 10\%$. \\ \\
{\it{gravitational lensing}}: Finally we come up with the relative error of  the gravitational lensing. This error can be estimated by the theoretical convergence introduce in Eq.(\ref{eq:kg}) using the standard $\Lambda$CDM cosmology with cosmological parameter $\Omega_m = 0.297 $, $H_0 = 0.704 $ and $\sigma_8 0.8$.
\begin{figure}
%\centering
\includegraphics[width=0.5\textwidth]{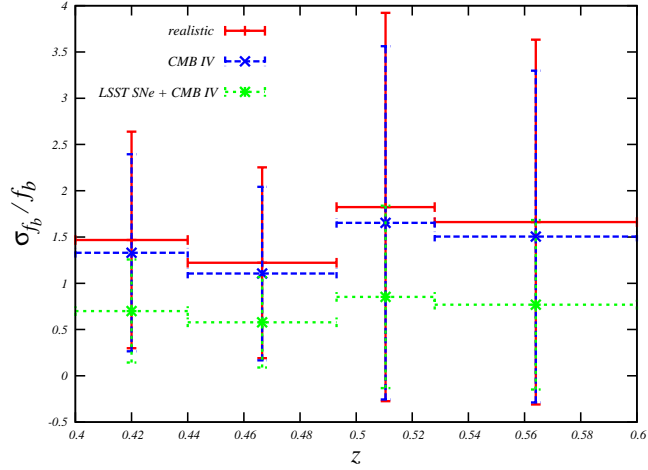}
\caption{ The relative error of the baryon fraction is plotted for different redshift bins for the realistic case (red solid line).  and the optimistic errors for CMB stage IV (blue dashed line) and CMB stage IV and LSST like SNe Ia data (green dotted line) }
\label{fig:rev-error}
\end{figure}
In order to show the effect of different type of the errors on baryon fraction measurement, in Fig.(\ref{fig:rev-error}) we plot the relative error with respect to the redshift. The redshift range of $(0.4,0.6)$ is divided to four bins with almost equal number of SNe Ia. The mean and the $1\sigma$ variance of the relative error is plotted (red solid line) for the realistic case that we considered in previous section. Then in order to study the effect of different error budget, we study the case where relative error in kSZ measurement can be improved by a factor of ten (by e.g CMB stage IV experiments) (blue dashed line). In the next step we plot the relative error estimation by assuming the characteristic resolution of LSST project which can improve the magnitude measurement to $0.2 mag$. As it is shown in Fig.(\ref{fig:rev-error}) there is much improvement in the relative error due to luminosity.
% also the GAIA project which can improve the current status of the luminosity measurement in local universe $z<0.15$ by an order of magnitude $\tilde{\sigma}_{sn (opt)} = 0.1 \tilde{\sigma} _{sn(r)}$ (pink dotted data points).
In order to conclude this subsection, we assert that the main contribution of the error comes from the SNe Ia measurement. The future experiments of SNeIa can improve the distance measurement drastically, which can bring this method in a new and practical state.
%However this procedure must determine the line of sight size of the cluster and the fraction of ionization.
%In order to estimate the accuracy of this method we use a Monte-Carlo simulated data for estimation of $\frac{\Delta T}{\bar{T}}|^{i}_{kSZ}$, by setting a mean value for the signal with $\Delta T = +1.92 \mu $ Kelvin, which is the mean observed temperature change in $218 GHz$ of the Atacama Cosmology telescope \cite{Hand:2012ui}. Then with a Gaussian distribution for a $\frac{\Delta T}{\bar{T}}|^{i}_{kSZ}$  with a $1\sigma$ error bar deviation in the mean value $\sigma_{\Delta T} = \pm 0.22 \mu $ Kelvin we let the error propagate in baryon fraction measurement.
In the next subsection we will discuss a very important issue of the probability of the detection of the configuration (SNe Ia in a BCG) which we proposed in this work.

 \subsection{The expected number of SNe Ia in a galaxy cluster}
In order to finalize our proposal for obtaining the baryon fraction, we should address the important question: "How many events of SNe Type Ia explosions are happened in the central galaxy of a cluster".
Accordingly in this subsection, we propose two distinct scenarios for the number of expected events for SNe happened in galaxy cluster.
First we anticipate that the SNe Ia explode in the central galaxy of a cluster. Then we assume that more than one SNe Ia happens in galaxy cluster.
%\subsubsection{SNe in a bright central galaxy of a cluster}
For the first scenario, we formulate this probability due to the simplified relation below where $N^{sn}_{obs}$ is the number of observed BCG with the desired condition of our proposal.
\be \label{eq:nobs}
N^{sn}_{obs} (T_{obs}; s; bcg ; \oslash )= f_{sky} \times N_{bcg} \times P_{\oslash} \times T_{obs},
\ee
where  $N^{sn}_{obs} = N^{sn}_{obs} (T_{obs};s;  bcg ; \oslash )$ is a function of the observation time $T_{obs}$ and also it depends on the cosmological volume of the survey and the portion of the sky $f_{sky}$ it spans. All of this information is encapsulated in symbolic parameter of $s$ which stands for survey. It is obvious that the observed number of configuration depends on physics of BCG ( indicated by "bcg" ) and the physics of SNIa (indicated by "$\oslash$" in eq.(\ref{eq:nobs})).
The $N_{bcg}$ is the number of BCGs that resides in clusters which has a dark matter host of mass $M$ with lower bound $ M_l < M _{bcg} $ , where $M_l$  indicate lower limit of a typical BCG  which resides in the redshift range of $z_i$ up to $z_f$. The parameter
$P_{\oslash}$ is the probability that a SNIa occurred in a BCG of a cluster in a year.
The number of BCG in galaxy clusters in a cosmological volume, which is limited by the redshift range, that we are interested in is as below
\be
N _{bcg} (M_l , z_i, z_f) = f_{bcg}\int _{M_l}dM \int_{z_i}^{z_f} dz \frac{dV}{dz} \frac{dn(M,z)}{dM},
\ee
where we set  $M_l = 10^{13.5} M_{\odot}$ as the lower limit of dark matter halo, the host of the BCG, $z_i = 0.4$ and $z_f=0.6$ is the lower and upper limit of the redshift survey that we are interested and $f_{bcg}$ is the fraction of luminous massive galaxies that resides in a center of galaxy cluster.
Due to the fact that BCGs are assembled  mainly by major mergers in galaxy formation and evolution process, there is an indication that the most massive galaxies must reside in galaxy clusters, accordingly we set $f_{bcg} \sim 1$ for this study.
In order to obtain the number density of cluster, we use the Press-Schechter approach to find $dn/dM$, specifically when we use the fitting function introduced in Jenkins et al \cite{Jenkins:2000bv}. Accordingly the number of BCG galaxies become $N_{bcg}  \simeq 5\times 10^5 \times f_{bcg} \times ({V_c}/{1 Gpc ^3})  $ , where $V_c$ is the cosmological volume. Note that the effective volume  is $V_c \approx 30 Gpc^3$ for  a survey which is in search of galaxies in redshift span of $0.4<z<0.6$.
Another important parameter is $P_{\oslash}$ to estimate is the  rate of SN Ia in a galaxy with the given mass range. In  Graur et al. \cite{Graur:2014bua}, there is an extensive study on the rate of Type Ia supernova. Graur et al. used  SNe samples to measure mass-normalized SNe rates as a function of stellar mass of the host galaxy and the star formation rate. By assuming the stellar mass of $10^{11.5}<M_*< 10^{12}$ (with the assumption of a mass to light ratio of $\sim10$) and the star formation rate we will have the marginalized fitting function as $P_{\oslash} \simeq 0.06 \times (M_* / 10^{12} M_{\odot}) / year$. It is worth to mention that the SN rate is proportional to star formation rate and specific start formation rate, accordingly the rate of SN decrease with evolution of BCGs from active to passive ones.
%However in low redshift ($z<0.15$), the specific star formation rate in BCGs is declining more slowly with time than for field or cluster galaxies, most likely due to the fuel from the cooling of inter-cluster medium \cite{McDonald:2015kym}.
To finalize this part, by considering all the complicated physics which is governing the SNIa rate relation with the host galaxy, we set very conservative rate of $P_{\oslash} \sim 0.006 / year$. \\
A project like Large Synoptic Survey Telescope (LSST) which is designed to operate for 10 years starting from 2019 and will capable of spanning the $20,000$ square degree of the sky (which means $f_{sky}\sim0.5$) in 6 optical bandwidth with a limited magnitude to a total point-source depth of $r \sim27.5$ \cite{Abell:2009aa}, we can estimate the  $N^{sn}_{obs} \sim 45000$ per year with the conservative assumptions we made in this section.
However we should note that one should also take into account that precise spectra for each of the SNe are absolutely necessary. This means that the SNe detection does not suffice, but a spectroscopic follow-up program should follow. Assuming that we will have a follow up of $10\%$ of the SNe Ia in a very conservative point of view in LSST life time project we can estimate the baryon fraction of  $\sim 45000$ clusters of galaxies.
%\subsubsection{Many SNe in a cluster of galaxy}
As a final word, the other scenario that can be used as a proposal for using the SNe for estimating the baryon fraction is that in a cluster of galaxy one can  monitor and measure all SNe in all galaxies within cluster, then {\it{on average}} the peculiar velocities of those SNe would be close to the bulk flow.
In this case, the error on the bulk velocity measurement can be decreased due to the number of SNe Ia found in a group/cluster of galaxies.
In the next section we will conclude the paper by future prospects.
%%//////////////////////////////////////////////////////////////////%%

\section{Conclusion and Future Prospects}
\label{Sec5}
The distribution of the baryons in the Universe is one of the main questions in cosmology, the big bang nucleosynthesis and cosmic microwave background radiation independently fix the baryon fraction. However the accessability of  baryons in late time is a challenging task and it seems that there is a missing baryon problem out there. The galaxy clusters as the main reservoir of baryonic matter which are filled with ionized electrons are suitable environments to study the distribution and physics of baryons in the Universe. One of the promising venues to address the distribution of baryons in the sky is the study of thermal and kinetic Sunyaev-Zeldovich effect which are used as a probe of ionized gas in clusters. In this work we propose a new idea/method to measure the baryon fraction using the kSZ effect and the SNe-Type Ia as the standard candle. For this method to work we assume that a SNIa explodes in a BCG. This is essential in a sense that the peculiar velocity of a BCG in a cluster can be used as an almost  fair representative of the cluster's bulk flow. The BCG resides in the depth of gravitational potential of cluster and its velocity with respect to the center of mass of the system is almost zero. \\ However in this work we indicate that the main uncertainty of velocity measurement comes from the systematics of SNe Ia including its progenitor velocity.
Accordingly we show that the SNe Ia in low redshift can be used to estimate the peculiar velocity of host galaxy.
In the other hand The kSZ signal depends on the baryon fraction and the bulk velocity of galaxy cluster. We assert that the deviation of a standard candle distance modulus from background prediction of the $\Lambda$CDM can be related to the peculiar velocity of SNIa host galaxy in lower redshifts, keeping in mind that the host galaxy is chosen to be a BCG.  We showed that by knowledge of the peculiar velocity and the temperature change of CMB we can constrain the baryon fraction. We investigate the Fisher forecast for the fiducial value of baryon fraction in the realistic (current) case and also in optimistic state. The analysis are discussed in the second subsection of Sec.(\ref{Sec4}).
In the observational prospect part, we also study the possibility of the observation of this effect. We estimate that in a future large scale survey, like LSST which spans half of the sky in $\sim 10$ years, we can do an optical spectroscopic follow up for almost $\sim 45000$ SNe Ia which explodes in a BCG in a cluster of galaxy. Also we should note the CMB stage IV experiments bring the number of individual cluster which can be examined for kSZ signal will grow gradually.
It is worth to mention that in the case of more statistics in galaxy cluster we can average the peculiar velocity of each host galaxy of SNIa , the average of squared velocity will be a representative of the bulk flow and the dispersion of velocities represent the error bar on bulk velocity.
%The story of the galaxy clusters with a BCG that host a SNIa can have another twist. In the case if we can pin down the properties of cluster like baryon fraction, the scale of a cluster and..., then we can find the expanding rate of the Universe via the proposal, that we made in this work encapsulated in  $\tilde {\kappa_v} $ which is related to physics we investigate here as: $\tilde {\kappa_v} = C^i f^i_b  \left[ (\frac{\Delta T}{\bar{T}}|^{i}_{kSZ}) / {\left(10 ^ {\Delta \mu /5} -1 \right)}\right]^{-1}$.
As a final remark we want to insist that future SNe Ia surveys will decrease the errors on baryon fraction and will bring the introduced method as a practically viable proposal.

\acknowledgments
We would like to thank Farhang Habibi for detailed and insightful comments on the manuscript. Also we would like to thank Saeed Ansari, Alireza Hojati, Nima Khosravi, Sohrab Rahvar, Sina Taamoli and Saeed Tavasoli  for useful discussions.
We thank the anonymous referee for his/her thorough review and highly appreciate the comments and suggestions, which significantly contributed to improving the quality of the manuscript.
SB acknowledges the hospitality of the Abdus Salam International Centre for Theoretical Physics (ICTP) during the final stage of this work.

\end{document}